\documentclass[cits]{JINST}
\usepackage{upgreek} 
\usepackage{url}
\usepackage{multirow}

\providecommand{\e}{\'{e} }
\providecommand{\etal}{ et al.\ }

\title{NEST: A Comprehensive Model for Scintillation Yield in Liquid Xenon}

\author{M. Szydagis$^a$\thanks{Corresponding
author.}, N. Barry$^a$, K. Kazkaz$^b$, J. Mock$^a$, D. Stolp$^a$, M. Sweany$^{a,b}$, M. Tripathi$^a$, S. Uvarov$^a$, N. Walsh$^a$, M. Woods$^a$\\
\llap{$^a$}University of California, Davis,\\
 One Shields Ave., Davis, CA 95616, USA\\
\llap{$^b$}Lawrence Livermore National Laboratory,\\
 7000 East Ave., Livermore, CA 94550, USA\\
 E-mail: \email{mmszydagis@ucdavis.edu}}

\abstract{A comprehensive model for explaining scintillation yield in liquid xenon is introduced. We unify various definitions of work function which abound in the literature and incorporate all available data on electron recoil scintillation yield. This results in a better understanding of electron recoil, and facilitates an improved description of nuclear recoil. An incident gamma energy range of $O$(1 keV) to $O$(1 MeV) and electric fields between 0 and $O$(10 kV/cm) are incorporated into this heuristic model. We show results from a Geant4 implementation, but because the model has a few free parameters, implementation in any simulation package should be simple. We use a quasi-empirical approach, with an objective of improving detector calibrations and performance verification. The model will aid in the design and optimization of future detectors. This model is also easy to extend to other noble elements. In this paper we lay the foundation for an exhaustive simulation code which we call NEST (Noble Element Simulation Technique).}


\begin{document}

\section{Introduction and Motivations} 
\label{sec:intro} 

Liquid noble elements (LNE) have been established as an attractive detection medium in experiments searching for dark matter or neutrino-less double-beta decay: a large number of existing or planned experiments employ liquid xenon, argon, or neon~\cite{Schnee2011,AprileDoke}. The direct dark matter search experiments looking for WIMPs (Weakly Interactive Massive Particles) aim to maximize their ability to discriminate between electron and nuclear recoils (ER and NR respectively) within their target volume. While the advantages of LNEs are abundant, one drawback is the non-linear energy dependence of the scintillation yield per unit energy deposited in the medium, for both ER and NR~\cite{Ni2006,Doke2002}. Further, this phenomenon is influenced by the magnitude of the applied electric field~\cite{AprileDoke}. In this article, we have addressed these issues for ER in liquid xenon by developing a general model for scintillation and ionization yields. We have implemented this model in a Geant4~\cite{Agostinelli2003} simulation and have demonstrated its ability to reproduce a wide variety of measurements, both for the field-free case and for high applied field.  We have also extended our model to include NR scintillation yield in the zero-field case.

An understanding of light yield is especially important in the low-energy region of interest ($\sim$5-50 keV) which contains the majority of the spectrum for NR from WIMPs in xenon (\cite{AprileDoke} and ref. therein). The energy dependence of light yield is non-monotonic at low energies, stemming from effects related to $dE/dx$, incident particle type, and electric field. This poses a challenge for simulations that employ constant scintillation yields. This model provides a comprehensive Monte Carlo similar to that of Tawara\etal\cite{Tawara2000} for NaI(Tl). Although this paper focuses solely on liquid xenon, in Section~\ref{sec:conclusion} we point to simple steps needed to include other noble elements.

\begin{figure}
\begin{centering}
\includegraphics[width=.75\textwidth]{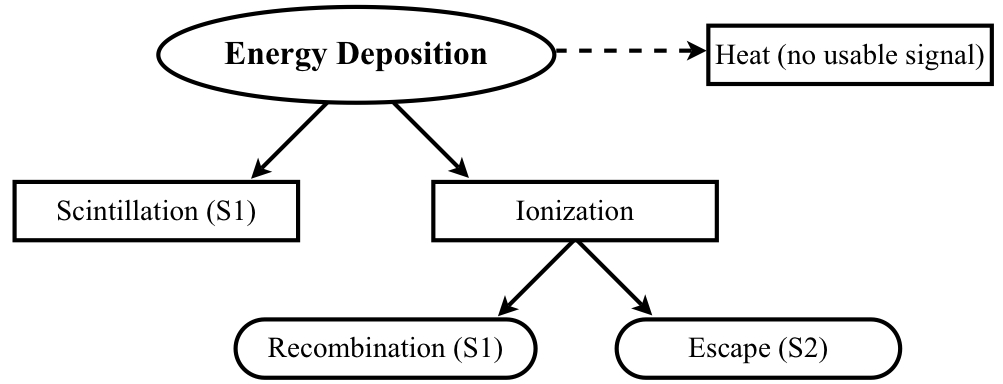} 
\caption[]{General schematic for the ionization, scintillation, and loss processes occurring after energy deposition by any recoiling species in a noble element. Ionization leads to electrons either recombining or escaping (as described in the text). Recombination forms excimers which contribute more scintillation, essentially identical to that generated by direct electronic excitation (collectively, S1). Escape electrons can be harnessed with an electric field and their number determined with a charge amplifier coupled to an anode, or, in a two-phase detector (see Section~\ref{sec:framework}), these electrons can produce more scintillation in a higher-electric-field gas phase (S2).}
\label{FlowChart}
\end{centering}
\end{figure}

\section{Physical and Mathematical Framework} 
\label{sec:framework} 

We begin by reviewing the basic principles of scintillation physics. A particle depositing energy creates excitation and ionization, as well as heat. Initial excitation contributes to the production of light called the primary scintillation, or S1, through an intricate process involving excited molecular states (unnecessary to fully simulate to reproduce the light yield)~\cite{AprileBook,Manzur2010,Doke1990}. The partitioning of deposited energy into excitation and ionization depends on incident particle type and possibly on electric field~\cite{Dahl2009}, but previous work indicates that the partition is independent of incident particle energy~\cite{Dahl2009,Doke1976}. This entire process is summarized in Figure~\ref{FlowChart}. Some energy is lost to heat in the form of secondary nuclear recoils~\cite{Lindhard1963,Sorensen2011} or sub-excitation electrons, which fail to lead to the generation of S1 light~\cite{AprileBook}. For ER, the former is negligible and the effect of the latter can be absorbed into a higher value for the work function (below).

Ionization electrons either recombine and lead to additional (S1) scintillation or escape. Two-phase LNE direct dark matter detectors {\cite{AprileDoke} drift such electrons with an electric field and extract them into a gas volume where they drift at higher velocity and generate more scintillation through more excitation and recombination in the gas. This is termed secondary, or S2, scintillation. The ratio of S2 to S1 light differs for ER and NR (\cite{AprileDoke, AprileBook} and ref. therein), and hence, it is the most critical discriminating parameter in dark matter searches.  The first step in fully understanding this phenomenon is to have a comprehensive description of ER behavior in xenon.

The distribution of the ionization electron population between recombination and escape constitutes the crux of the energy dependence of the S1 yield.  The direct excitation contribution to S1, which depends on the small ratio of excitons to ions,  is consequently small (\cite{Dahl2009} and ref. therein). Since an ionization electron lost to recombination cannot escape and one which escapes cannot recombine, S1 and S2 light yields are anti-correlated. Furthermore, scintillation yield, and hence recombination probability, is known to be a function of the energy loss per unit of path length $dE/dx$~\cite{Doke2002} and of the applied electric field~\cite{Dahl2009}.

For the energy partition we use a simplified Platzman equation~\cite{AprileDoke,AprileBook}
\begin{equation}
E_{dep} = N_{ex} W_{ex} + N_i W_i = N_i (\alpha W_{ex} + W_i),
\quad \alpha \equiv N_{ex} / N_i
\label{eqn2.1}
\end{equation}
where $E_{dep}$ is energy deposited by a particle in a single interaction (not its total energy deposition), $N_{ex}$ is the number of excitons created per deposition, $N_i$ the number of ions, $W_{ex}$ the work function (required energy) for exciting atoms, and $W_i$ the ionization work function. One sums the numbers of excitons and ions stemming from single interactions (recoils) to calculate the total yield of one incident particle. The ratio of excitons to ions, often labeled as $\alpha$, may differ for NR versus ER, and may change with field, but it is likely not a function of energy~\cite{Dahl2009}. The theoretical value of this dimensionless ratio $\alpha$ is a constant 0.06 for liquid xenon~\cite{Takahashi1975,Miyajima1974}, used successfully to fit experimental data~\cite{Manalaysay2010}. Some measurements of $\alpha$ in liquid xenon have been as high as 0.20, though arguably consistent with 0.06 within experimental uncertainties~\cite{Doke2002,Aprile2007}.

Ignoring heat, including the kinetic energy of ionization electrons (which can be absorbed into $W_{i}$), and assuming near-100\% efficiency for excited or recombined electrons to lead to S1 (well-established experimentally~\cite{Dahl2009}), the numbers of quanta produced are
\begin{eqnarray}
N_i &=& \frac{E_{dep}}{\alpha W_{ex} + W_i} \quad \mbox{and} \\
N_{ex} &=& \alpha N_i
\label{eqn2.2and2.3}
\end{eqnarray}
which in turn lead to $N_{ph}$ photons and $N_{e}$ ionization electrons:
\begin{eqnarray}
N_{ph} &=& N_{ex} + r N_i \quad \mbox{and} \\
\label{eqn2.4}
N_e &=& N_i (1-r)
\label{eqn2.5}
\end{eqnarray}
where $r$ is the recombination probability. For long particle tracks we use~\cite{Ni2006,Doke1988}
\begin{equation}
r = \frac{A \frac{dE}{dx} }{1 + B \frac{dE}{dx}} + C, \quad C = 1 - A/B,
\label{eqn2.6}
\end{equation}
derived from Birks' Law~\cite{Birks1964}. Doke\etal\cite{Doke1988} derived a scintillation yield for LNEs from Birks' Law, while we adopt the same approach but instead derive the recombination probability underlying the scintillation yield by separating excitation from the recombination probability. The first term of the equation represents the recombination which occurs when a wandering ionization electron is captured by an ion other than its parent (termed ``volume'' recombination). The second term ($C$) represents ``geminate'' recombination, also known as Onsager recombination~\cite{Doke1988,ThomasImel}. It is defined as ionization electrons recombining with their parent ions, which is assigned a fixed probability for all $dE/dx$. A low $\alpha$ implies that most S1 light is a result of the recombination processes.

We plot Equation \ref{eqn2.6} in Figure~\ref{RecombProb0} with the parameter values that we will show in Section \ref{sec:zero_field} to reproduce the light yield of gamma rays, the first particles we consider in this work. The constraint is imposed so that as $dE/dx$ approaches infinity, the recombination probability approaches unity.\footnote{Linear energy transfer LET, defined as the quotient of $dE/dx$ and density, is often used in place of $dE/dx$.} Even in the absence of an electric field, some ionization electrons do not recombine on timescales of interest to experiments (<1 ms); they temporarily escape the Coulomb fields of the ionized xenon atoms and are called ``escape'' electrons~\cite{Doke1988}. For instance, $\sim$10\% of ionization electrons from a 122 keV gamma in xenon will not be observed to produce excitons~\cite{Dahl2009,Bolozdynya2008,Aprile2006}.

\begin{figure}
\begin{centering}
\includegraphics[width=.70\textwidth]{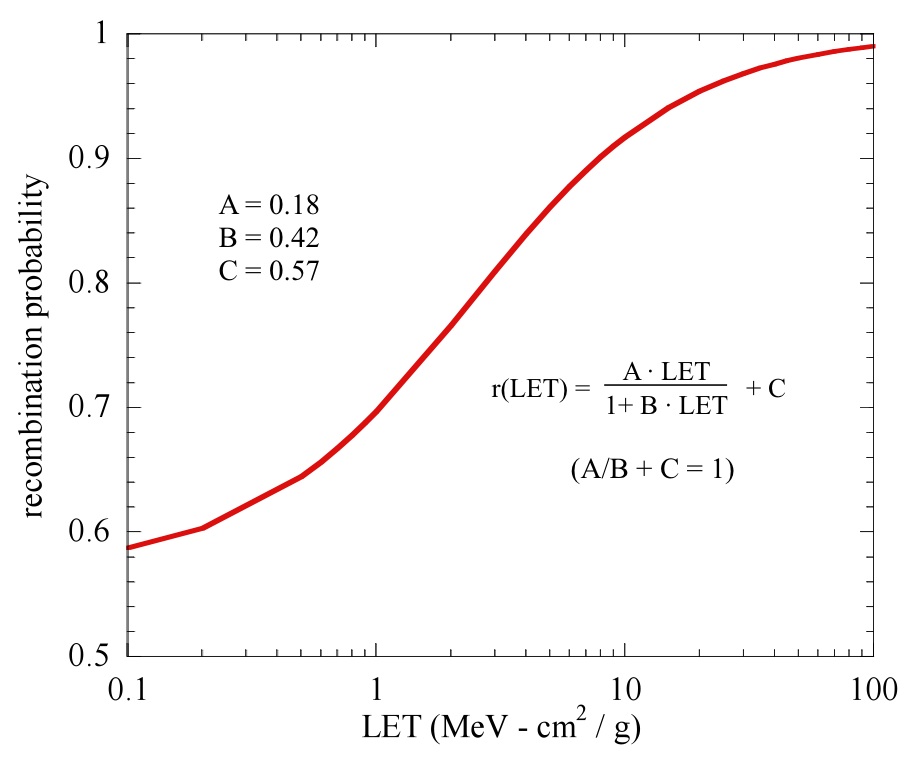} 
\caption[]{The electron recombination probability that is implemented in our Monte Carlo simulation for zero electric field. It is a function of Linear Energy Transfer (LET), which is the quotient of $dE/dx$ and density. We base this probability on Doke et al.\ \cite{Doke1988}, treating $A$, $B$, and $C$ as free parameters and using experimental data on light yield to constrain them.  The parameters compare well with past values used such as $C=0.55$~\cite{Ni2006}. The parameters are also constrained by the condition $A/B+C = 1$, which ensures complete recombination at infinite LET, or $dE/dx$.}
\label{RecombProb0}
\end{centering}
\end{figure}

For short particle tracks the Thomas-Imel box model is better suited~\cite{Sorensen2011}, where the recombination probability is instead~\cite{Dahl2009,Sorensen2011,ThomasImel}
\begin{equation}
r = 1-\frac{\mbox{ln}(1+\xi)}{\xi},\quad \xi \equiv \frac{N_{i}\alpha'}{4a^{2}v}
\label{eqn2.7}
\end{equation}
where $\alpha'$ is a constant dependent on ionization electron and hole mobilities and the dielectric constant (beyond the scope of this work), $v$ is the mean ionization electron velocity, and $a$ is a length scale defining ionization density volume. We can equate $\alpha'/(a^{2}v)$ with $e^{2}/(a\epsilon kT)$~\cite{Dahl2009}, where $e$ is the electron charge, $\epsilon$ is the dielectric constant, $k$ is Boltzmann's constant, and $T$ is the electron temperature. Using $\epsilon=1.96$, $T=165$ K~\cite{Conti2010}, and $a=4.6~\upmu$m~\cite{Mozumder1995}, we estimate a value of 0.14 for zero electric field. We assume here that most electrons will thermalize prior to recombination (and for $T$ use the 1 atm boiling point, near which most detectors operate~\cite{AprileDoke}). This is false at high electric field, which increases electron energy~\cite{Dahl2009}.

We define ``short tracks'' to be those that are shorter than the mean ionization electron-ion thermalization distance, 4.6 $\upmu$m in liquid xenon~\cite{Mozumder1995}. If an incident gamma or electron produces secondary gammas or scatters multiple times, then there are multiple interaction sites. At each site, tracks are separately summed to determine whether to apply Thomas-Imel or Doke/Birks formalism. Each is based on the same recombination physics, albeit with different descriptions governed by short- or long-track limits~\cite{Dahlpriv}. The Thomas-Imel understanding relies on recombination in a ``box'' (cubic) geometry~\cite{Dahl2009,ThomasImel} while Doke constructs his formulation utilizing the cross-section of a long column of electron-ion pairs~\cite{Doke1988}. We seamlessly integrate these two models into one unified picture of scintillation yield, which is continuous versus energy and consistent with data.

Our model determines recombination for individual recoils in a Geant4 simulation~\cite{Agostinelli2003} by explicitly tracking energies and recoil ranges for secondary electrons produced by an incident gamma. Doing so preserves macroscopic characteristics of the energy-dependent light yield (Figure~\ref{MoneyPlot}) while remaining stochastic. As demonstrated in Figure~\ref{Histograms}, Monte Carlo fluctuations in tracks result in a variation in the number of scintillation photons ultimately produced, even before fluctuation in the ratio of excitons to ions, or their sum (Fano factor), is taken into consideration.\footnote{The latter effect is well-known to be small, sub-Poissonian in fact~\cite{AprileDoke,Dahl2009,Seguinot1995}.}

\begin{figure}
\begin{centering}
\includegraphics[width=.999\textwidth]{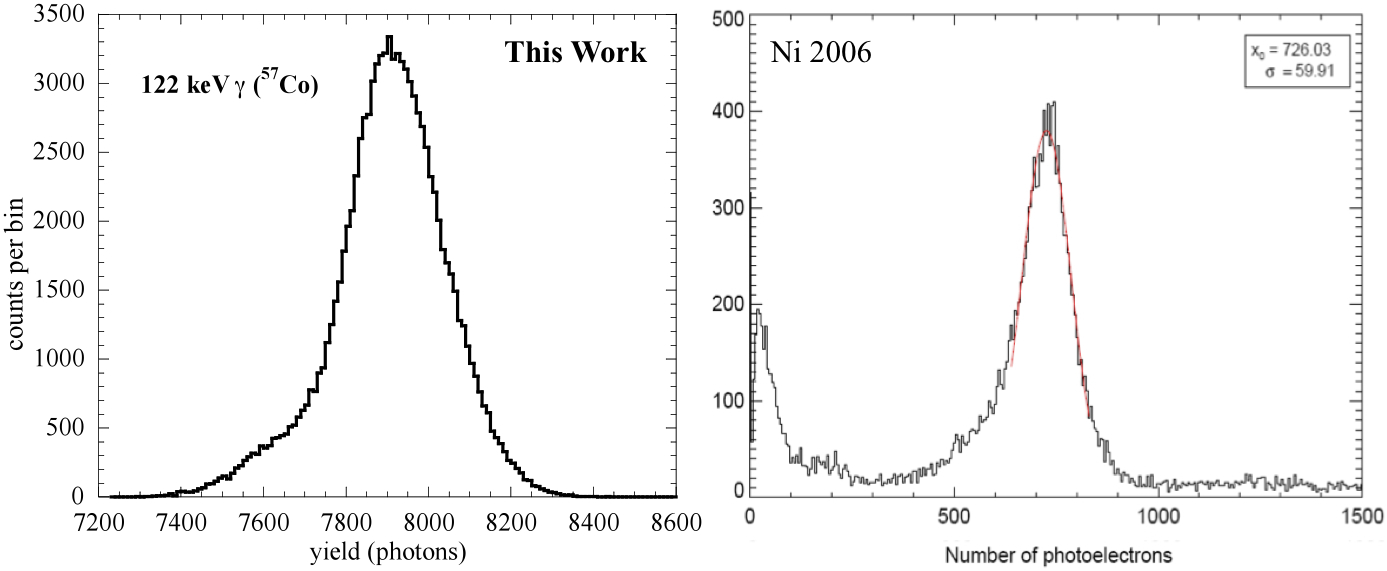} 
\caption[]{(left) A histogram of photon yields obtained in NEST in response to mono-energetic 122 keV gammas ($^{57}$Co). The width of the response curve demonstrates the ability of our model to generate variation in photon yield, even when no standard means of scintillation fluctuation is included, such as that described by the Fano factor~\cite{Seguinot1995}, or detector effects. Even at this minimal level, our simulation produces non-Gaussianity in the low-energy tail. (right) Markedly similar real data~\cite{Ni2006}. Compare with the asymmetry of our simulated peak. The reason for observing spread in final light yield, sans Fano factor and detector effects, lies in the dependence of the recombination probability on a detailed track history and stochastic variations in $dE/dx$ values.  Because of $\sim$10\% photon detection efficiency, our x-axis (photons) differs by a factor of ten (right axis in photoelectrons).  Also, the low-energy peak, in the data on the right, comes from partial energy deposition, which is not possible in our simulated infinite xenon volume.}
\label{Histograms}
\end{centering}
\end{figure}

For simplicity, our model defines a mean work function for production of either excitons or electron-ion pairs, such that $E_{dep} = W(N_{ex}+N_{i})$. This is equally as effective as defining two distinct work functions for successfully explaining experimental results~\cite{Doke2002,Dahl2009,Shutt2007}. Different values for the two processes are difficult to extract experimentally, and using a mean value causes no loss of generality~\cite{Dahl2009,Dahlpriv}. Because $W$ is also equivalent to $(\alpha W_{ex}+W_{i})/(1+\alpha)$ based on Equations \ref{eqn2.1} to \ref{eqn2.2and2.3}, a higher $W_{ex}$ is canceled out by a lower $W_{i}$ or vice versa, leaving the total numbers of both quanta unchanged. Similarly, a higher $\alpha$ is counteracted in the above equation by a lower $W_{ex}$, making an exact experimental knowledge of $\alpha$ less relevant.

In general, $dE/dx$ increases with decreasing energy for electrons less than 1 MeV~\cite{ESTAR}. Hence, recombination probability (and scintillation yield) should increase with decreasing energy, as long as the Doke model applies~\cite{Doke2002, Doke1988}. However, an interesting phenomenon occurs in a low-energy region where the scintillation yield instead decreases (Figure~\ref{OboAndNaI}). For gamma rays below $O$(10~keV), the slope of the light yield curve changes sign~\cite{Obodovskii1994}, as it does for NaI(Tl) in a different energy regime~\cite{Doke2008,Murray1961}, albeit due to a depletion of activation sites in the crystal~\cite{Murray1961}.

\begin{figure}
\begin{centering}
\includegraphics[width=.6\textwidth]{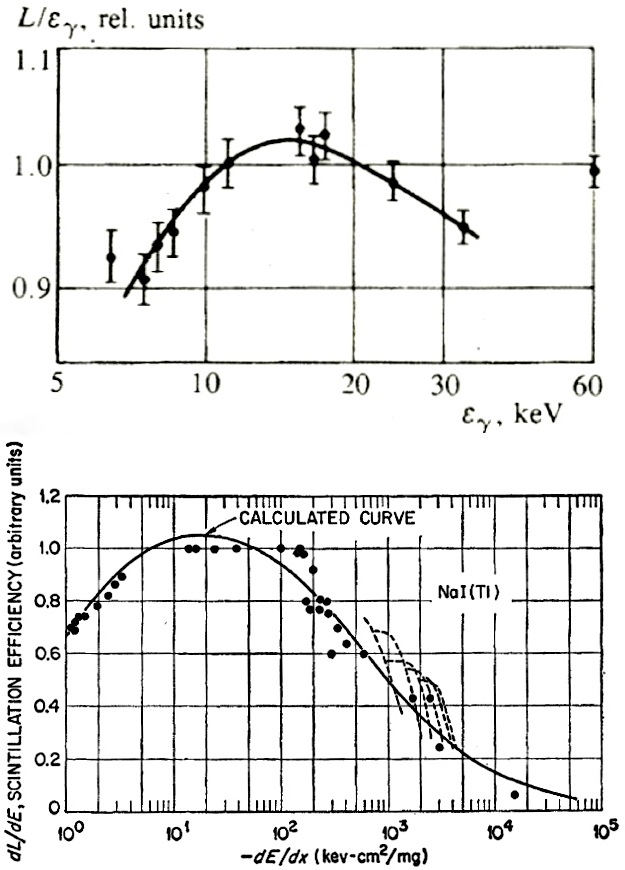} 
\caption{(top) This figure is re-printed from~\cite{Obodovskii1994}. It shows the anomalous decrease in scintillation per unit energy with increasing $dE/dx$ (decreasing gamma energy) observed in liquid xenon for the lowest energy gammas studied. (bottom) A demonstration of a similar high-$dE/dx$ scintillation yield decrease for electron recoils in NaI(Tl). Here the decrease is caused by a well-known mechanism involving the kinematics of interactions within the crystal. This figure is re-printed from~\cite{Murray1961}. Note that the x-axis of the top figure is energy, while for the bottom it is $dE/dx$.}
\label{OboAndNaI}
\end{centering}
\end{figure}

One would expect recombination to be enhanced at high $dE/dx$ due to greater ionization density, but it has been shown that at low energies (high $dE/dx$) recombination becomes independent of ionization density~\cite{Dahl2009,Sorensen2011,ThomasImel}. This is accommodated in our model because low-energy particles are strictly in the Thomas-Imel regime, where the recombination probability depends on the energy, via the number of ions, and not on $dE/dx$. In other words, for particle tracks smaller than the thermalization distance of ionization electrons, the total track length is no longer relevant. Dahl was the first to apply a modified Thomas-Imel box model to successfully explain the odd turnover in light yield at low energy, which changes position in energy for different electric fields~\cite{Dahl2009}.  At higher energies, the Thomas-Imel box model had already been successfully utilized to fit data on the global scale, such as total charge yield (total number of ionization electrons extracted) versus electric field, glossing over microphysics~\cite{Manalaysay2010}. In Dahl's approach, the model is applied to the scale of the lowest energy (sub-keV and eV) individual electron recoils using PENELOPE~\cite{Salvat2001}.  

In our model, we apply the Thomas-Imel model to the electron-recoil energy scale $O$(1 keV) accessible with Geant4, making the assumption that it can be adapted for zero electric field.  Manalaysay et al.~\cite{Manalaysay2010} demonstrate that the Thomas-Imel parameterization, even though undefined at zero field, can have a finite limit and produce realistic results. Moreover, Figure~\ref{MoneyPlot} in this work demonstrates the utility of this assumption by successfully reproducing light yields.

\section{Zero-Field Scintillation Yield vs.\ Energy} 
\label{sec:zero_field} 

The parameters of Equation \ref{eqn2.6} and $\alpha'/(4 a^{2}v)$ of \ref{eqn2.7} (henceforth $\xi/N_{i}$) were treated as free in order to match our compilation of existing data on zero-field gamma ray scintillation.  A concerted effort was made to include all known experimental results in our compilation and any possible omitted works are unknown to the authors rather than intentionally excluded. Figure~\ref{MoneyPlot} compares the results from simulations for this work to the available empirical data.

A traditional $\chi^2$ analysis with errors could not be performed because  many papers do not quote error bars in either their figures or their text (e.g.,~\cite{Barabanov1987}), or state that their errors are negligible~\cite{Ni2007}. Furthermore, there are several instances of more than one data point for a given gamma energy, with one or more points lacking errors, and yields are most often quoted as relative to measurements made at 122, 511, or 662 keV. The latter is due to experimental issues, such as difficulties in knowing PMT quantum efficiency at low temperatures and UV wavelengths and/or inadequate knowledge of the reflectivity of the walls of the detector. Thus, assumptions had to be made in order to translate all results into absolute yields. We used 511 and 122 keV as benchmarks, translating uses of 662 keV or other energies by experiments as benchmarks as needed. For the low-energy data of Obodovskii and Ospanov~\cite{Obodovskii1994}, the 59.5 keV $^{241}$Am line is used.

The process of translating the data into absolute yield increases the error by an amount which can not accurately be estimated due to the varying age and quality of the data, and in some rare cases, due to variation in choice of benchmark energy, which has a profound effect. For instance, if the results of Barabanov\etal\cite{Barabanov1987} are re-scaled assuming a benchmark of 122 keV in Figure~\ref{MoneyPlot}, then 59.5 and 122 keV are not outliers, but all other points become too low. Choices like this needed to be made to bring the majority of a large quantity of old and new data into rough agreement. Overall, the ultimate consistency among different detectors is remarkable. Precedence for such comparison of different data sets is exemplified by Doke\etal\cite{Doke2002}, who were forced to assume a match in yield between one $^{22}$Na gamma line and a high-energy electron in order to incorporate the results of Barabanov\etal into their Figure 4. We treat data all as equally as possible in Figure~\ref{MoneyPlot} without averaging results except when necessary, such as when we calculated a mean 122 keV yield in order to incorporate data which referenced it but not 511 keV (which was our primary benchmark because it was as common as 122 keV in experiments, but provided the additional advantage of forcing the majority of data from Barabanov\etal to agree with other works).

\begin{figure}[h]
\begin{centering}
\includegraphics[width=.934\textwidth]{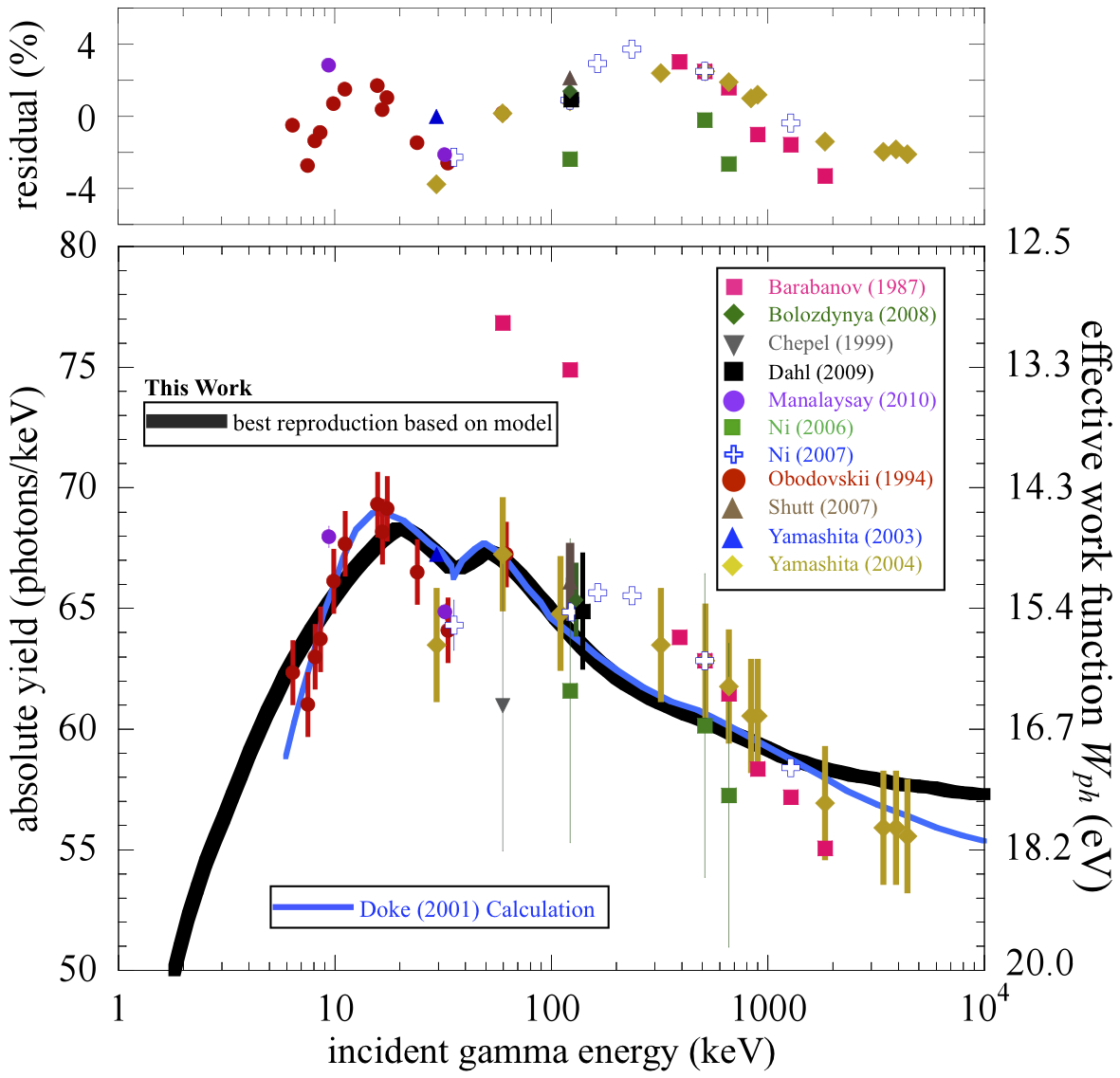} 
\caption[]{Experimental values for absolute S1 yields for ER in xenon as a function of incident gamma energy compared with our Monte Carlo output, the foundation of which is the recombination probability curve in Figure \ref{RecombProb0}. Fitted Gaussian means (see Figure~\ref{Histograms}, left) are used to report the simulated scintillation yields. Data are taken from~\cite{Ni2006,Dahl2009,Manalaysay2010,Bolozdynya2008,Shutt2007,Obodovskii1994,Barabanov1987,Ni2007,Chepel1999,Yamashita2004,Yamashita2003}. When authors quote relative yields, we infer absolute numbers of photons using the 511, 122, or 59.5 keV line as a reference energy.  All data are translated based on the works where absolute yields are available~\cite{Ni2006,Dahl2009,Bolozdynya2008,Shutt2007}. The systematic error within one detector should be comparable at each energy, making relative measurements meaningful. The cluster of data points at 122~keV have been separated for clarity. For these and other overlapping points, some error bars are thicker than others for differentiation. The 3-5 MeV points are from a preprint of~\cite{Yamashita2004} and are reproduced here with permission of M. Yamashita for completeness, though are considered less reliable due to poorer resolution. The 9.4 and 32.1~keV de-excitations of $^{83m}$Kr \cite{Manalaysay2010} are not comprised solely of gamma rays, but are also included in the figure for completeness. Two points, at 164 and 236 keV (hollow blue crosses \cite{Ni2007}), are from inelastic neutron scatters, resulting in gammas of these energies together with a nuclear recoil component, which may possibly serve to increase the yield. The right-hand y-axis uses the definition $W_{ph} = E_{dep}/N_{ph}$, as described in the text. More features and significant outliers are addressed in Section~\ref{sec:zero_field}, where the work of Doke\etal\cite{Doke2001}, which is in good agreement with our own, is also explained in depth.}
\label{MoneyPlot}
\end{centering}
\end{figure}

\begin{table}[t]
\centering
\caption[]{The parameters of our model that best reproduce the data. Note that the value of $\xi/N_{i}$ is comparable to our earlier estimate of 0.14. The minimum mean-squared residual for the high energy data alone, lying primarily in the Doke/Birks regime, was calculated in two different ways: treating multiple points for the same energy independently, or averaging all the points at the same energy. Since some data sets are without stated uncertainty, the latter was a raw, not error-weighted, average and thus less reliable. The parameters resulting in the best reproduction are the same in either case.}
 \begin{tabular}{|@{} lc|c|r @{}}
 \hline
       & data window and other conditions & parameters for best reproduction (this work) \\
\hline
       & low energy points alone ($<$15 keV) & $\xi/N_{i} = 0.19$ \\
       & high energy points treated independently & $A = 0.18, B = 0.42, C = 0 .57$ \\
       & averaging points (raw) at same energy & $A = 0.18, B = 0.42, C = 0.57$ \\
\hline
\end{tabular}
\label{chisq}
\end{table}

We minimize the mean-squared of the residuals in order to optimize the parameters of our model, which is founded on Equations \ref{eqn2.6} and \ref{eqn2.7}. We perform a series of full Monte Carlo simulations with different sets of parameters until finding the optimal set. Then our model can successfully reproduce the available data, thus predicting yield in regions where the data is sparse. The two clear outliers from Barabanov\etal\cite{Barabanov1987} are excluded. They may be explained by non-uniform light collection in the detector, as claimed by~\cite{Yamashita2004}. Also discarded, for the purpose of our plot and the optimization, was one result considerably distant from all others, quoting less than 34~photons/keV at 22 and 88 keV~\cite{Belli1993}. Below 15 keV, the Thomas-Imel box model dominates, and it is in this regime where we find the optimal value for $\xi/N_{i}$. We then use that value for the data as a whole, keeping it fixed. The top of Figure~\ref{MoneyPlot} is a plot of the percent residuals between various data sets and our model. Given our assumptions for relative-absolute translation, most data do not differ by more than $\pm$4\% from a simulation with the parameters which best reproduce the data. These are listed in Table~\ref{chisq}, above.  Throughout this work,  we do not fit data on scintillation yield directly. Instead we systematically explore ansatz solutions for the recombination probability which are then entered into the simulation to reproduce some measured yields and predict others.

In addition to the list in Table~\ref{chisq}, other parameters which we employed, but that we held fixed, included $W = 13.7$ eV, $\alpha = 0.06$, and 4.6 $\upmu$m as the electron-ion thermalization length scale, which we pragmatically selected to determine the cross-over energy where the Doke model would begin to dominate, because of how well it predicts the approximate location of the major change in slope in scintillation yield as a function of energy. The $W$-value is an error-weighted average from reports by Doke\etal\cite {Doke2002} and Dahl~\cite{Dahl2009}: 13.8 $\pm$ 0.9 and 13.7 $\pm$ 0.2 eV. At least three other values exist, but were not included in the average: 14.0 eV because it was reported without a clearly stated uncertainty~\cite{Aprile2010a}, 13.46 $\pm$ 0.29 eV~\cite{Shutt2007} due to a calibration issue expressed by Dahl~\cite{Dahlpriv} (but it was still used to estimate a 122~keV yield), and 14.7 eV~\cite{Doke1990,Miyajima1974,Tanaka2001}, superseded in the past thirty-five years by more measurements benefiting from technological improvements.

The y-axis labeled ``effective'' work function\footnote{This work function is also known as $W'_{ph}$~\cite{Doke1990} or $W_{s}$~\cite{Belli1993}, conflicting with others' use of the same nomenclature~\cite{Seguinot1995}.} in Figure~\ref{MoneyPlot} is for $W_{ph} = E_{dep}/N_{ph}$, which depends upon incident energy, because $N_{ph}$ varies with energy via the recombination probability $r$ (Equation \ref{eqn2.4}). This definition expedites comparison with other works (\cite{Chepel1999,Doke1999} and ref. therein) which do not relate all results to one $W$. The ratio of $W$ to $W_{ph}$ is described as scintillation efficiency\footnote{This efficiency should not be confused with the light collection efficiency of a detector.} and the relationship $W_{ph} \geq W$ must hold true~\cite{Ni2006}. Equality implies excitation alone, with no ionization. The inverse of our $W = 13.7$ eV corresponds to 73 photons/keV.

The clear dip in light yield at $\sim$30-35 keV is present because of a resonance at the xenon K-edge for production of low-energy Auger electrons~\cite{Yamashita2004,Yamashitapriv,Tojo1985,Obodovskii1993}. They are low enough in energy ($\leq10$ keV) to be governed by the Thomas-Imel model. Furthermore, at or above about 29.8 keV, the K-shell x-ray is emitted, traveling a small distance to create a second interaction site for which $r$ is separately calculated, serving to enhance the effect of the scintillation decrease: tracks are summed separately at each site to determine which recombination treatment to use, enhancing the probability of being in the Thomas-Imel regime. We note that the simulated dip is too shallow, possibly due to an inability of Geant4 to produce electrons of low enough energy to accurately depict the necessary interactions, even with the Auger physics module activated. In Geant4, a gamma does not generate electrons whose energy is so low that their range would be below the minimum distance cut-off. In order to maintain energy conservation, an energy deposition is recorded for the gamma in the tracking output file. To partially mitigate this issue, we assume that an ER with that recorded energy occurs~\cite{Dahlpriv}.


Thus far we have only discussed gamma rays. We now treat electrons as the incident particles and subject them to a similar simulation as was done for incident gammas. We obtain $W_{ph} = 19.4 - 14.7$~eV, for incident electron energies in the range 1 - 100 keV.  For electrons below $O$(100~keV) in energy, S\'{e}guinot\etal report $W_{ph} = 14.3$ eV~\cite{Seguinot1995} in agreement ($<$5\%) with the lower edge of our predicted range. Given the energy dependence of $W_{ph}$, S\'{e}guinot et al.'s measurement must be an effective value, most likely an average or a minimum. In their earlier work~\cite{Seguinot1992}, they claim a value of 12.5 or 12.7 eV, in disagreement with their later work and our present work, but Chepel\etal\cite{Chepel1999} cite S\'{e}guinot\etal as 12.7 $\pm$ 1.3 eV, where the upper bound on the uncertainty (14.0~eV) disagrees by only $\sim$5\% with our prediction (14.7 eV). We cannot, however, reconcile their claim of a low ionization work function (defined later in this section) for low-energy electrons of 9.76~eV~\cite{Seguinot1995,Seguinot1992} using any of the possible definitions. Further, it has been disputed~\cite{Doke1999,Miyajima1995,SeguinotReply}.

Unlike for low-energy electrons, data on high-energy ones are plentiful~\cite{Ni2006}, derived from $^{207}$Bi decay, which is a mixture of 0.976 and 1.048 MeV electrons, and a 1.064~MeV gamma, plus other lower-energy decay products~\cite{Aprile1991}. Authors typically quote yield as that of an approximate ``1~MeV electron,'' implying that the electrons dominate; cited yields are all in agreement with each other within uncertainties~\cite{Ni2006, Doke2002}. We take two typical empirical results to compare to our simulation: 42.0 $\pm$ 8.4~photons/keV~\cite{Doke1990} and 46.4$^{+2.2}_{-4.3}$~photons/keV.\footnote{0.64 $\pm$ 0.03 relative yield with respect to $W = 13.8$ eV~\cite{Doke2002}.} The model as it stands predicts too high a yield, 57.4 photons/keV for a 1 MeV electron. As an aside, for a 1 MeV gamma the simulated yield of our model is 59.2 photons/keV. This difference in yield is explained by the lower recombination probability for the ionizations from the ER of an incident 1 MeV electron compared to the lower-energy (higher-$dE/dx$) ER produced by a (Compton scattering) 1 MeV gamma ray. However, this difference is too small to explain the much lower yield in 1 MeV electron data.

Our initial attempt to match the absolute yield of 1 MeV electrons using the Doke model was not successful. Forcing a match results in too low of an absolute yield for other well-established results. Setting the (relative) 122~keV point from Barabanov\etal\cite{Barabanov1987} equal to one of the absolute measurements of the 122~keV yield lowers the yield of the 1 MeV gamma, hence that of the 1 MeV electron as well, creating better agreement with data. Unfortunately, this makes other Barabanov\etal points too low, in direct conflict with much more recent measurements \cite{Ni2007,Yamashita2004} and barely within the quoted error of Ni\etal\cite{Ni2006}. (Ni\etal report absolute, not relative, yields, requiring no assumptions in interpreting their results.) The problem at hand is subtly evident in the work of Doke\etal (the basis for the curve in Figure \ref{MoneyPlot} labeled Doke 2001). They use data from Yamashita\etal\cite{Yamashita2004} at the expense of ignoring Barabanov\etal\cite{Barabanov1987} and determine the absolute yields of 1 MeV electrons and gammas to be nearly identical \cite{Doke2001}; using 122 or 511 keV as the key to translate relative into absolute yield makes the absolute yield too high again for a 1 MeV electron as in our own work. In Ni\etal\cite{Ni2006}, one possible solution can be found: using a universal (average) $dE/dx$ for interactions (explained in Section \ref{sec:other_models}), the Doke model can be made to fit the 1 MeV electron yield along with equal or lower-energy gammas, but we wish to avoid losing the stochastic nature of the simulation. Finally, even if one were to try applying an existing model other than Doke's, the similarity in $dE/dx$ between a gamma and an electron at 1 MeV and the direct contradiction between one older set and two or three newer data sets would remain points of contention.

\begin{table}[t]
\centering
\caption[]{The predictions of our model for the S1 photon yield of a 1 MeV electron at zero electric field, under various changes in parameters. Our model can reproduce the reported experimental yield to high precision, but only under extreme conditions, fully explained below in the text.}
 \begin{tabular}{|@{} ll|cr @{}|}
 \hline
 
       & conditions (parameters) & yield (ph/keV) & \\
\hline
       & $A = 0.18, C = 0.57, \alpha = 0.06$ (unchanged model parameters) & 57.4 & \\
       & $A = 0.00, C = 0.57, \alpha = 0.06$ (geminate/Onsager recombination alone) & 43.1 & \\
       & $A = 0.00, C = 0.57, \alpha = 0.20$ (additional initial excitation) & 46.7 & \\
       & most recent experimental data, with the smallest uncertainty~\cite{Doke2002} & 46.4 & \\
\hline
\end{tabular}
\label{bismuth}
\end{table}

An extreme solution to the challenge is to assume that non-geminate recombination ceases for a minimally ionizing particle. Parameter $A$ of Equation \ref{eqn2.6} represents such recombination, and is proportional to ionization density, which is in turn proportional to $dE/dx$~\cite{Doke1988}. Perhaps low ionization density prevents ionization electrons from recombining with non-parent ions because of distance, though Mozumder claims that this is impossible \cite{Mozumder1995}. The thermalization distance (4.6 $\upmu$m) is much greater than the reach of the Coulomb field of a xenon ion as defined by the Onsager radius (49 nm)~\cite{Doke2002}, so Onsager recombination should not be dominant. For the sake of argument, we surmise that a thermalized ionization electron can eventually be re-attracted to its parent ion if no other ion is available. Alternatively, while thermalizing, an ionization electron could become trapped in a closed path within the Coulomb field of the parent ion if repelled by neighboring electron clouds~\cite{Kubota1979}. This may help explain the slow recombination time observed by Doke and Masuda~\cite{Doke1999}. However, we do not present our solution as a definitive one, given a lack of fundamental plausibility, but instead as a pragmatic one, allowing the creation of a simulation which produces reasonable results for gammas and electrons alike in a wide range of energies. A summary of our results for the 1 MeV electron under parameter variation is presented in Table~\ref{bismuth}. Changing $\alpha$ to correspond with the results of~\cite{Doke2002} in addition to setting $A = 0$ matches data best. A higher $\alpha$ may imply more Onsager recombination than encoded in $C$. More such recombination is hard to distinguish from a shift in the initial ratio of excitons to ions~\cite{Dahl2009}.

After reproducing zero-field scintillation yield it is possible to make predictions of measurements other than yield. Aprile\etal report that the zero-field escape fraction for ionization electrons from a 662 keV gamma is $\chi = 0.22 \pm 0.02$~\cite{Aprile2007} (also called $\beta$~\cite{Manzur2010}). From our simulation we derive $\chi = 0.19$. With an $\alpha$ of 0.20 and the adjustment of $A = 0$ explained above, we can reproduce the $\chi = 0.43$ reported for $^{207}$Bi decay~\cite{Doke2002}. In addition to $\chi$ we can derive a value for the effective ionization work function, defined as $W_{i}^{eff} = E_{dep}/N_{i}$. The definition of $W_{i}^{eff} = W (1+\alpha)$ is equivalent~\cite{Doke2002,Doke1990,Tanaka2001}.\footnote{Just like the other work functions, the notation is inconsistent. This one is variably called simply $W$~\cite{AprileDoke, AprileBook}, or $W_{q}$~\cite{Dahl2009}. Our $W$ is at times denoted $W_{ph}(max)$~\cite{Doke2002}, $W_{sc}$~\cite{AprileBook}, $W_{s}$~\cite{Seguinot1995}, or $W'$~\cite{Doke1988,Doke1999}.} Our model indicates that it should be 14.5 eV, in agreement with the 14.2 $\pm$ 0.3 eV reported by Obodovskii and Pokachalov (as cited by~\cite{AprileBook}), but not with the oft-cited 15.6 $\pm$ 0.3 eV (\cite{Doke1990,Aprile2006} and ref.\ therein). The discrepancy is easily explained: 15.6 eV corresponds with $\alpha = 0.14$ if $W = 13.7$ eV, or $\alpha = 0.06$ coupled with the outdated value of $W=14.7$ eV \cite{Miyajima1974}. One more result we have now is our simulated ratio of $W_{i}^{eff}$ to the band gap energy in liquid xenon of 1.56, in good agreement with the theoretical (1.65) and measured ($\sim$1.6) values~\cite{AprileDoke}. We are continuing to collect more data to compare with the model, making improvements to the model whenever possible.

\section{Discussion of Other Models} 
\label{sec:other_models} 

In this section, we discuss the advantages of our model as compared to others that are available in the literature. Figure~\ref{MoneyPlot} includes the calculation of Doke\etal\cite{Doke2001}. This approach is based on extracting an electron response from a portion of the gamma data~\cite{Obodovskii1994,Yamashita2004},  applying that extracted response to early-generation secondary electrons and then reconstructing the gamma response. In contrast, our model, which uses Equations \ref{eqn2.6} and \ref{eqn2.7}, is applied to each new electron created in the cascade, as low in energy as permitted by Geant4. The range of each electron is obtained from Geant4 output, and not assumed from a Bethe-Bloch calculation. Doing so allows for variations in $dE/dx$.  Such variations contribute to the spread in scintillation yield (Figures~\ref{bismuth} and \ref{Histograms}). This approach can be incorporated in any simulation which outputs lists of energy depositions, and utilized for predicting charge yields (Section~\ref{sec:quenching_with_field}) and the NR quenching factor (Section~\ref{sec:nuc_recoil}). Unlike some other previous works~\cite{Ni2006,Doke2002}, we do not utilize a direct fit to the energy dependence of total scintillation yield. We thus predict, not extrapolate, the response to an energy regime that is lower than what has been studied experimentally.

In earlier work, similar models have been applied to materials other than LNE, for instance, NaI(Tl), which is a well known scintillator. The recombination probability ($r$) for NaI(Tl) has the same functional form~\cite{Murray1961} as Equation \ref{eqn2.6}, albeit with $C = 0$, leaving $A = B$. Despite the compositional differences between mono-atomic LNE and solid crystal scintillators (often comprised of multiple elements, including dopants), or organic scintillators for which Birks' Law was derived originally, this formula has been applied to LNE with modest success~\cite{Ni2006,Doke2002,Doke1988}. However, its application involved defining a universal $dE/dx$ for a gamma interaction, which had to be obtained by averaging over all interactions (ERs). Electron numbers and energies of course vary~\cite{Tawara2000}, diminishing the utility of a mean $dE/dx$ by leading to a static recombination probability. Our treatment of xenon avoids usage of such an average.

There was, however, no certainty that Equation \ref{eqn2.6}, which works relatively well with global $dE/dx$, would also work when accounting for individual $dE/dx$. We explored numerous other possibilities with the same qualitative features (low probability at low $dE/dx$ and high probability at high $dE/dx$), but these were unmotivated and made for poor reproductions of the data ($erf(x)$, exponential, Equation \ref{eqn2.6} with $dE/dx$ raised to different powers, and non-unity $r$ at infinite $dE/dx$.) No benefit was seen in inventing a new approach, and using $C = 0$ (no geminate recombination) did not minimize the mean-squared error, while a non-zero $C$ did. Numerous other physically-motivated models exist, Mozumder's for example \cite{Mozumder1995}, but have more free parameters, are inapplicable at zero field, with no simple adaptation, or apply only to MeV energies \cite{Dahlpriv}. We achieve agreement with the most data and the fewest parameters. There also appeared to be no benefit to the introduction of heat loss for ER, to explain the trend of low-energy gamma data. The $dE/dx$ of a low-energy ER is comparable to that of an NR~\cite{Aprile2006}, but the Lindhard mechanism, by which NR lose energy to additional nuclear recoils, does not apply to low-energy gammas, which should exclusively produce ER~\cite{Dahl2009,Lindhard1963}.

Our model is ultimately an approximation because Geant4 does not accurately simulate electron production (tertiary, quaternary generations, etc.)~below 250~eV~\cite{Agostinelli2003}, unlike other software packages with more precise low-energy electron physics, such as PENELOPE~\cite{Dahl2009,Salvat2001}. However, we demonstrate that working within the default Geant4 energy regime is sufficient for all practical purposes. As shown in Figure~\ref{MoneyPlot} (top), our work can match data in most cases to better than 5\%, assuming an accurate conversion from relative to absolute gamma and electron yield at zero field. In addition, the model continues to work well for non-zero field, as shown in the next section. 

There are various extensions to the existing simulation packages which address scintillation directly. For example, RAT \cite{Gastler2010} is a Geant4 add-on that simulates the scintillation yield of many elements, with the exception of xenon. Our modeling of ionization and our focus on xenon, both lacking in RAT \cite{RAT2}, make our work complementary.

\section{Scintillation Quenching with Electric Field} 
\label{sec:quenching_with_field} 

As electric field strength increases, ionization electrons liberated by an interaction are increasingly less likely to recombine with an ion. This is known as electric field scintillation quenching (\cite{AprileDoke} and ref. therein), not to be confused with the quenching factor for NR, which involves different scintillation loss mechanisms such as the Lindhard effect~\cite{Manzur2010,Lindhard1963,Sorensen2011}. For a two-phase detector, this results in an increase in S2 light output at the expense of S1. The problem is quite complex because recombination probability does not change uniformly across all energies~\cite{Dahl2009}. Dahl demonstrates that a fundamental theory of recombination as a function of electric field is possible, which would explain the trend in recombination probability using a modified Thomas-Imel box model~\cite{Dahl2009}. For reasons delineated below, we approach the problem semi-empirically, adapting as free parameters $A$, $B$, and $C$ from Equation \ref{eqn2.6}, and $\xi/N_{i}$ from Equation \ref{eqn2.7}. We tune these on a subset of available data sets, and then accurately predict many others.

An alternative would be to attempt a first-principles approach in Geant4 by creating look-up tables for the recombination probability $r$ based on a Thomas-Imel application in PENELOPE. For three reasons, we decided against this approach. First, the introduction of tables would eliminate the benefits of the stochastic nature of simulations, an attribute that was a primary motivation for implementing our model. Even a look-up table with stochastic variations fails to capitalize on the details of track history captured by Geant4, thus making it less realistic.  Second, the first-principles approach breaks down at zero field, where single-phase detectors operate.  Third, it is not verified experimentally above $O$(100 keV) as opposed to the approach employed by Doke\etal\cite{Ni2006,Doke2002,Doke1988}.  Energies in the 0.5 - 3.0 MeV range are important for double-beta decay experiments and use of xenon in PET scans~\cite{Ama2009}. Thus, our hybrid model of Thomas-Imel plus Doke covers a large range in both electric field and energy.

 \begin{table}
\centering
\caption[]{Summary of the Thomas-Imel parameter $\xi/N_{i}$ in our model, as compared to Dahl (who does not discuss zero field). The different columns for Dahl are, respectively, ER only, both ER and NR, and ER and NR where $\alpha$ is field-dependent for NR. Our work is for ER and a constant $\alpha$. We maintain $\alpha = 0.06$ at all fields.}
 \begin{tabular}{|@{} lc|c|ccccr @{}|}
 \hline
 \multicolumn{8}{|c|}{$\xi/N_{i}$}\\
 \hline
       & Electric Field (V/cm)    & This Work & \multicolumn{4}{c}{Dahl 2009\cite{Dahl2009}}  & \\
\hline
       & 0 & 0.19 & \multirow{6}{*}{} & - & - & - & \\
       & 60 $\pm$ 5 & 0.034 & & 0.0339(2) & 0.0348(2) & 0.0386(2) & \\
       & 522 $\pm$ 23 & 0.029 && 0.0335(3) &  0.0354(2) & 0.0342(2) & \\
       & 876 $\pm$ 36 & 0.026 & &0.0296(3)  & 0.0303(2) & 0.0299(1) & \\
       & 1951 $\pm$ 86 & 0.023 && 0.0371(5) & 0.0348(3) & 0.0319(2) & \\
       & 4060 $\pm$ 190 & 0.021 && 0.0317(5) & 0.0318(4) & 0.0280(2) & \\
\hline
 \end{tabular}
 \label{dahl} 
 \end{table}

We began our data-matching with one data set that is the most comprehensive in terms of simultaneous span in energy (2-200 keV) and electric field (60-4060 V/cm)~\cite{Dahl2009}. By starting with one very complete data set, we initially sidestepped the issue related to collecting data sets from different detectors with distinct systematic effects. First, we let $\xi/N_{i}$ vary, using data below 15 keV to determine the best reproductions of data. Our results, in Table~\ref{dahl}, differ from the values used by Dahl~\cite{Dahl2009} for the same data; despite the disagreement the data is reproduced well, as demonstrated later in Figure \ref{DahlElectron}. Agreement with data in spite of disagreement in the value of a parameter from the same model  is likely a consequence of not applying the Thomas-Imel formula to the lowest energy ($< 250$ eV) electrons. Geant4 does not generate them, and we are forced to utilize the Thomas-Imel box model in an approximate fashion. (We adapted Thomas-Imel to zero field at low energy without an extrapolation, but with a recalculation, in Section~\ref{sec:zero_field}). Another contributing factor is the fact that Dahl applies the Thomas-Imel model completely to all his data, while we use it only at low energy, switching to Doke/Birks for the longer tracks of particles at higher energies. The lack of tracking of the shortest tracks causes the Thomas-Imel model to break down in Geant4 at high energy, but we take advantage of the regime where it does work. After having found values for $\xi/N_{i}$ at five different fields, we fit a power law to it: $\xi/N_{i} = 0.057 E^{-0.12}$ (field in V/cm). This is consistent with Dahl's expectation of $\xi/N_{i} \propto E^{-0.1}$ \cite{Dahl2009}.

The next step was to find the Doke parameters as a function of electric field magnitude. We began this study by looking at the well-established electric field dependence of the 122 keV gamma. The result for 122 keV has been repeated several times~\cite{Dahl2009,Aprile2006,Aprile2005}, with duplicate and triplicate points often taken at the same electric field to verify repeatability~\cite{Aprile2006}. We proceeded to check the variation of the Doke parameters achieved studying only this one energy against other energies and to adjust our fit for the electric field dependence of the parameters accordingly, though continuing to give 122 keV results the greatest weight.

There must be a transition between the Thomas-Imel regime and the Doke regime. At zero field we use the ionization electron thermalization length scale as the cross-over distance, and this immediately makes the transition smooth. As electric field increases a need arises to change this cross-over distance in order to avoid creating a discontinuity. We offer a power law fit as a function of electric field as a first effort without optimization: the distance in $\upmu$m is $69/\sqrt{E}$, where the electric field magnitude is in units of V/cm.

\begin{figure}[t]
\includegraphics[width=1.0\textwidth]{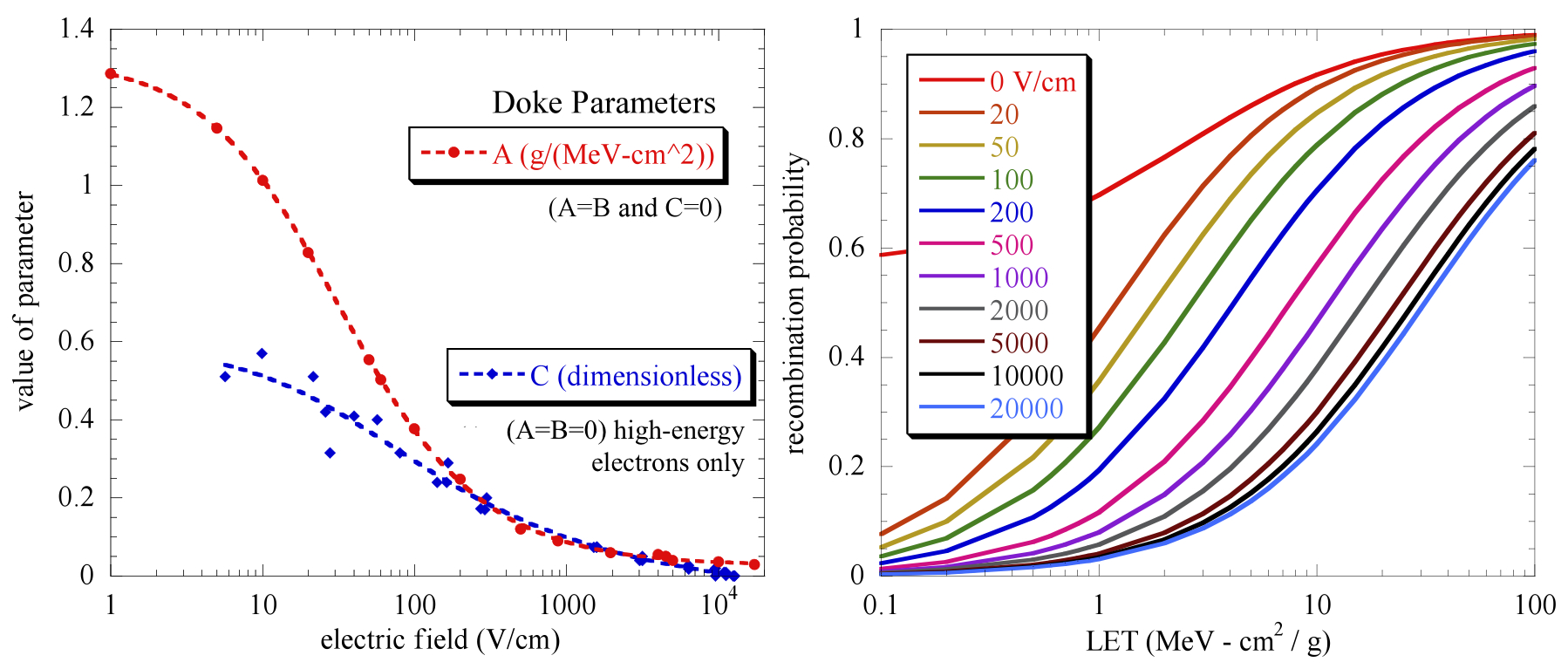}
\caption[]{(left) Choice of parameters ($A$, $B$, and $C$ from Equation \ref{eqn2.6}) and their behavior as a function of electric field. There are two distinct cases. The red curve was determined by matching available gamma data, giving the most weight to the 122 keV gamma. In the second case, the blue curve is for relativistic electrons, matched to 1 MeV electron data. We propagated our extreme assumption of $A = 0$, and then varied $C$. Decreasing values of  $A$ and $B$ imply decreasing volume recombination; a decrease in $C$  corresponds to a  decrease in Onsager recombination (see text). We use a power law fit to interpolate $A$ and $C$ between discrete fields. The final outcome is a decreasing recombination with increasing field, as expected. The case of zero field is a discontinuity where $A = 0.18$, $B = 0.42$ (both zero for high-energy electrons), and $C = 0.57$ (Figure~\ref{RecombProb0} and Section~\ref{sec:zero_field}). (right) Variations in recombination probability, as a function of applied electric field, used in our simulation. This set of curves is based only on using $C = 0$ and the variation in $A$~($=B$) as depicted at left. The high-energy electron case is not plotted because it is the trivial case of constant recombination probability with respect to LET.}
\label{ABCchange}
\end{figure}

\begin{figure}[!h]
\begin{centering}
\includegraphics[width=0.99\textwidth]{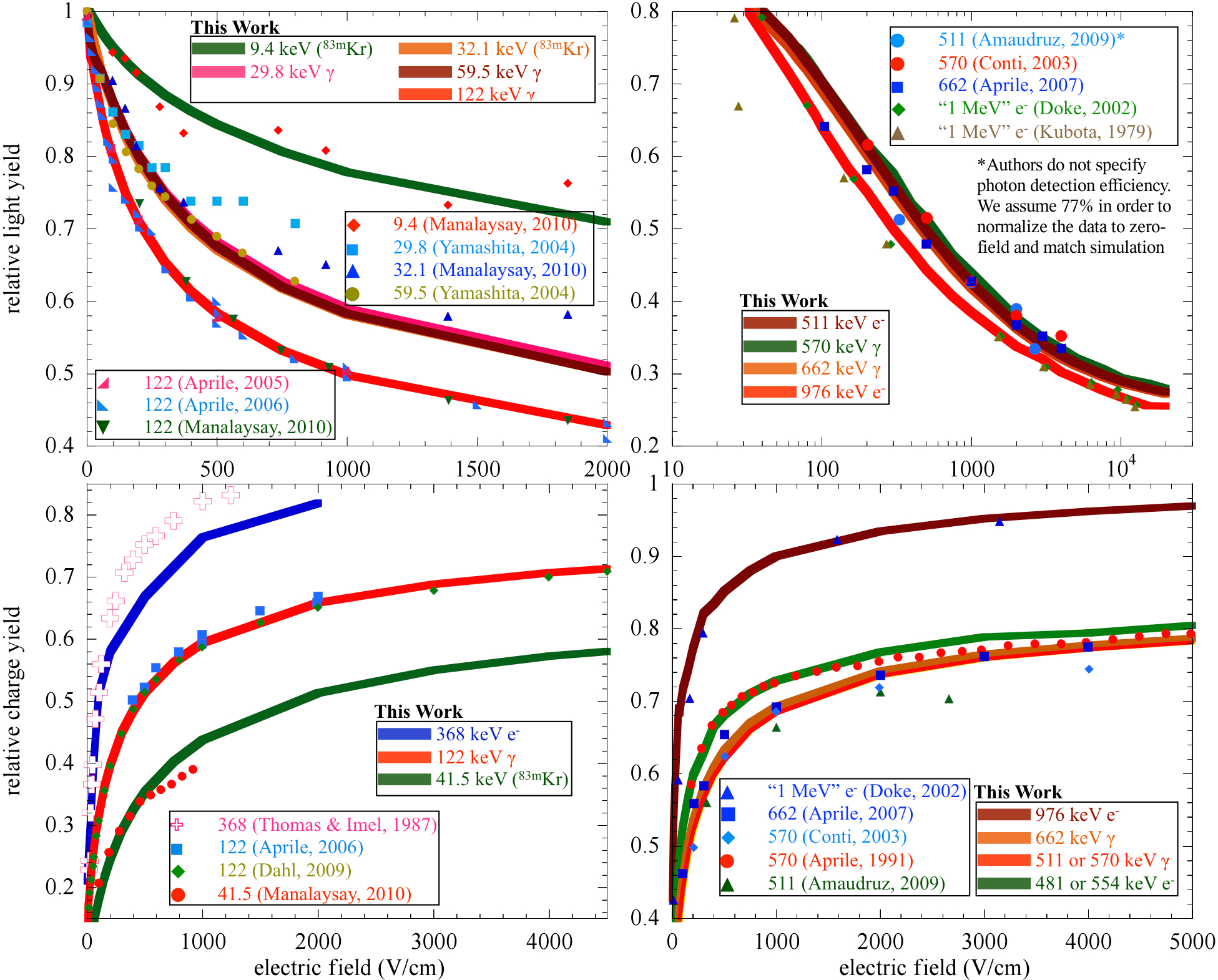} 
\caption[]{(top row) Comparison of our model with experimental data on relative scintillation yield as a function of electric field. The relative yields of 1.0 are unique to each energy and are obtained by normalizing to zero-field values.  In several plots our curves lie on top of each other if too similar in energy. Our model is valid to higher fields as well, but we restrict these plots to select ranges which are rich in data. Empirical data are from~\cite{Doke2002,Manalaysay2010,Aprile2007,Aprile2006,ThomasImel,Yamashita2004,Kubota1979,Aprile2005,Conti2003}, labeled by the first author in each case.

(bottom row) Comparison of our model with data on relative charge yield as a function of electric field, with data taken from~\cite{Doke2002,Dahl2009,Manalaysay2010,Aprile2007,Aprile2006,Aprile1991,Ama2009,Conti2003}. In this case, the relative yields of 1.0 are again unique to each energy, but correspond to the infinite-field ionization electron yield. This normalization assumes that at infinite field all possible charge is extracted. Thus the relative yield can equivalently be thought of as the escape fraction for ionization electrons separated from ions. ``All possible charge'' is defined as the number of ions or electrons generated \emph{ab initio}, assuming $\alpha = 0.06$ (0.20 for 1 MeV electrons). Experimental results quoted are appropriately re-scaled if assuming a different value of $\alpha$ either explicitly or implicitly ($W_i^{eff} = 15.6$ eV may imply a different $\alpha$ for example)~\cite{Aprile2007,Aprile2006}, as are ones plotting results in terms of the fraction of total quanta instead of the fraction of electrons~\cite{Dahl2009}. Two contradictory 570 keV gamma charge yield measurements are shown. Interestingly, the data set (Aprile 1991~\cite{Aprile1991}) that is inconsistent with our model matches well with the 554 and 481 keV electrons, which are associated decay products of the 570 keV gamma from $^{207}$Bi decay.  The green curve in the bottom right plot has been generated using this interpretation. Underestimated charge yield for the 368 keV electron (blue line compared to pink crosses) may be explained by a detector effect: ionization electron multiplication in charge collectors~\cite{Doke1999}.}
\label{ElectricField}
\end{centering}
\end{figure}

\begin{figure}[!h]
\begin{centering}
\includegraphics[width=0.9\textwidth]{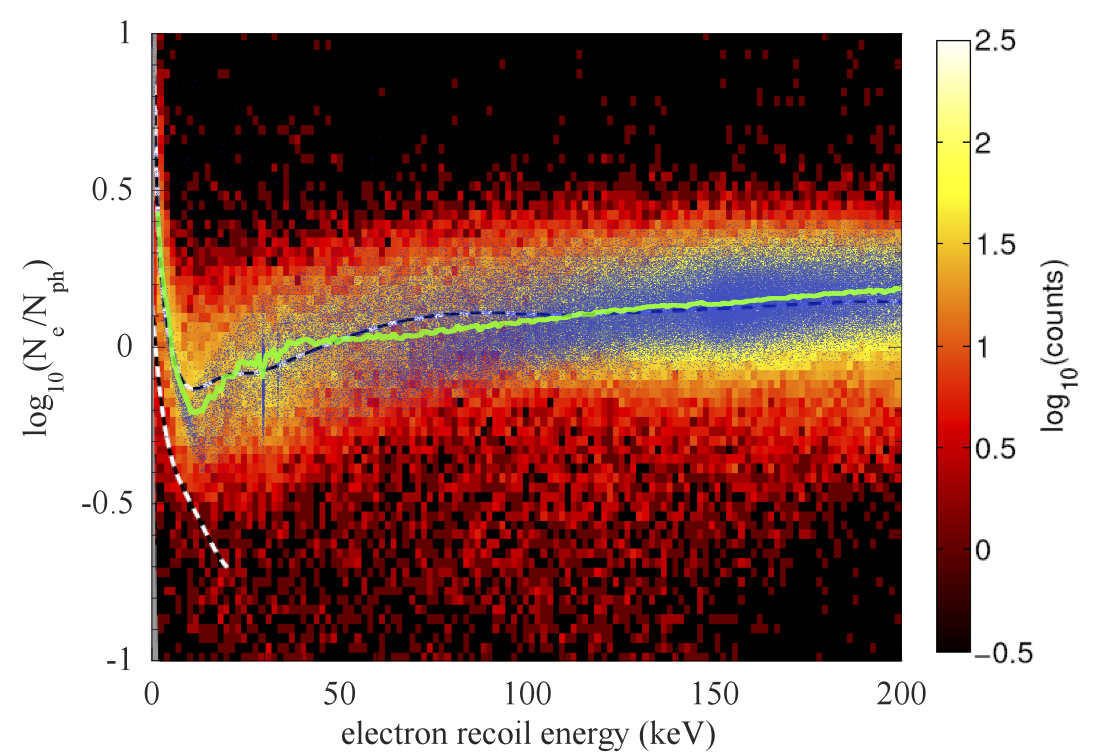}
\caption[]{The logarithm of the ratio of the number of ionization electrons (the basis of S2) to the number of S1 scintillation photons, generated by the ER stemming from a high-energy gamma ray (356 keV in both simulation and data), at an electric field of 876 V/cm. Actual data, from the Xed liquid xenon detector, is depicted as a color map (red to yellow scale), reproduced from Figure 5.1 of C.E.\ Dahl's Ph.D.\ Thesis~\cite{Dahl2009}. The output of our simulation model is superposed as a simple density plot in blue.  The green curve is a raw average of our simulated points in 1 keV bins. Vertical marks at $\sim$30 keV are likely related to the xenon K-edge, or are binning artifacts. Because we do not simulate the Xed geometry, we do not study and compare the density of our points to the density of the data as encoded by the coloration. To approximate the Xed result we look at the energy deposition in a finite volume of xenon of comparable dimensions to Xed but without walls. The scatter in the simulation is a result of recombination fluctuations  discussed in the text: particle history variation and $dE/dx$ fluctuations, which swamp Fano factor and $\alpha$ effects. The experimental data have additional scatter, either due to a low-yield tail in S2 or a high-yield tail in S1. The former arises from detector effects, wherein ionization electrons can be absorbed by walls~\cite{Dahl2009}. The upper dashed curve represents the centroid of the ER band as shown in the original figure~\cite{Dahl2009}, while the lower one is the NR band, beyond the scope of this work.}
\label{DahlElectron}
\end{centering}
\end{figure}

Monotonic variation emerged naturally in our model (Figure~\ref{ABCchange}) for $A$ and $B$, and a negligible value for $C$, so we set $C = 0$. These parameters have physical meaning as discussed earlier. We interpret a near-zero value for $C$ as demonstrating that volume recombination dominates over Onsager recombination. This corresponds exactly with the findings of Dahl, who reports that Onsager recombination should be negligible~\cite{Dahl2009}. With an external electric field overwhelming the local fields of ions, this is expected. In the case of zero field, the opposite appears true ($C = 0.57$), but this is empirically supported in argon~\cite{Doke1988}, with available evidence being ambiguous for xenon~\cite{Ni2006,Doke2002}. This is one approach, but we were unable to explain as much data as easily with a fully physically-motivated model for electric field, the Jaff\e model, $r = 1 - 1/(1+K/E)$, where $K$ is the so-called recombination coefficient and $E$ is the electric field. However, the Jaff\e model came close to matching significant amounts of data as modified by Aprile\etal to take into account delta electrons~\cite{AprileBook,Aprile1991}. Our approach to the electric field dependence has the advantage of adapting the existing formulae for zero field instead of introducing more models.

Our total compilation of comparisons is available in Figure~\ref{ElectricField}. The data points are shown without any error bars for either the yield or the electric field magnitude, for clarity. (There are few cases where they are even available.) A significant amount of recent data is predicted accurately, having begun by reproducing 122 keV points and the low-energy regime in Dahl's data. Discrepancies exist, but were inevitable, due to contradictory results being present in such a large compilation (likely due to non-obvious and unreported systematic errors), and due to the fact that we never directly fit any of these experimental yield-dependence curves, even at 122~keV. Instead, we reproduce or predict data with simulation by testing differing underlying recombination probability curves. With a few parameters that describe the recombination process for individual recoils and are functions of electric field, we can effectively explain global aspects such as total light and charge yields. These yields combine individual yields from all recoils spawned by one incident particle. This is a clear break from the past work of others who directly fit the total yields with models (for example, \cite{Aprile1991,Manalaysay2010,Dawson2005}).

We omit works (\cite{Dawson2005} for example) which define ``total'' light and charge yields for partial energy deposition ``peaks,'' as these would require proper modeling of the detector in question in order to reproduce well. We also exclude older works with measurements that cannot be reconciled with more recent ones, including \cite{Barabash1985}, and Voronova\etal as cited in~\cite{Dawson2005}. More recent experiments have benefited from technological advancements, leading to more accurate measurements. Recent 9.4 keV data is poorly predicted by the Thomas-Imel model. Curiously, agreements can be achieved by approximating the $^{83m}$Kr de-excitation as only a gamma ray and applying the Doke model despite the low energy. Without such adjustments these data contradict Dahl~\cite{Dahl2009} in this energy range. The contradiction may arise from inaccurate decomposition into constituent electrons and gammas in Geant4.

In spite of these few faults, we have successfully reproduced many measurements based on various detectors. Figure~\ref{DahlElectron} shows our prediction for $\log_{10}(N_{e}/N_{ph})$ as superposed on the data from Dahl \cite{Dahl2009}. While the low-energy regime is a reproduction based on these data, the extension above $\sim$15 keV is the prediction from our model. Here we have compared simulation with data at only one of Dahl's electric field choices, as an example. Comparable agreement is seen at other values of electric field as well.

\section{Nuclear Recoil Quenching Factor} 
\label{sec:nuc_recoil} 

We have applied the framework motivated in Section~\ref{sec:framework}, and validated in Section~\ref{sec:zero_field}, to zero-field NR scintillation yield. NR scintillation efficiency has been a subject of debate, with several contradictory experimental results and theoretical underpinnings~\cite{Aprile2010,Collar2010}. This is an issue of great significance for dark matter detectors. Dark matter particles are expected to produce nuclear, not electron, recoils in most theories~\cite{Schnee2011}.

We offer a unique approach here (applied earlier only by Dahl~\cite{Dahl2009}, but at non-zero field) that uses parameters from a model motivated by gamma-induced ER and applies them to NR directly. Tracks of NR below $O$(100 keV) are always within the Thomas-Imel regime~\cite{Dahl2009}. We again use the single free parameter from the Thomas-Imel model ($\xi/N_{i}$) that successfully explained the low-energy yield of ER. We make only one modification to the mathematical framework for NR, such that Equation \ref{eqn2.1} is modified~\cite{Dahl2009,Sorensen2011} as follows:
\begin{equation}
E_{dep}  L(E_{dep}) = (N_{ex}+N_i) W = (1+\alpha) N_i W
\label{eqn6.1}
\end{equation}
where $L$ is the Lindhard factor (or, effective Lindhard factor for modification to the Lindhard theory made to better explain data). It should not be confused with $\mathcal L_{eff}$, which represents the ratio of scintillation light produced by an NR to that produced by a 122 keV gamma, at zero electric field. Both are functions of energy, and the latter is the traditional way to report results on NR. Though not equivalent to $L$, knowledge of $\mathcal L_{eff}$ enables an empirical determination of $L$, which represents the fraction of energy available for excitation and ionization. Unlike ER, NR will lose most of its energy ($\sim$80\%) to heat, which in this case implies interactions with other nuclei instead of electrons. Nuclear interactions with electrons excite or ionize atoms and thus produce usable signals for a detector~\cite{AprileBook,Sorensen2011}. The fractional factor $L$ is the reason why two energy scales are often defined: keV$_{ee}$ and keV$_{nr}$ (or keV$_{r}$)~\cite{AprileDoke,Dahl2009}. The former is for ER and the latter for NR. In this paper we use only ``keV'' as the total energy of a recoil as reported by Geant4. Without knowing $L$ or $\mathcal L_{eff}$ one can not \emph{a~priori} define the NR energy scale in terms of the ``electron equivalent'' energy in ``keV$_{ee}$.''

\begin{figure}[!h]
\begin{centering}
\includegraphics[width=1.0\textwidth]{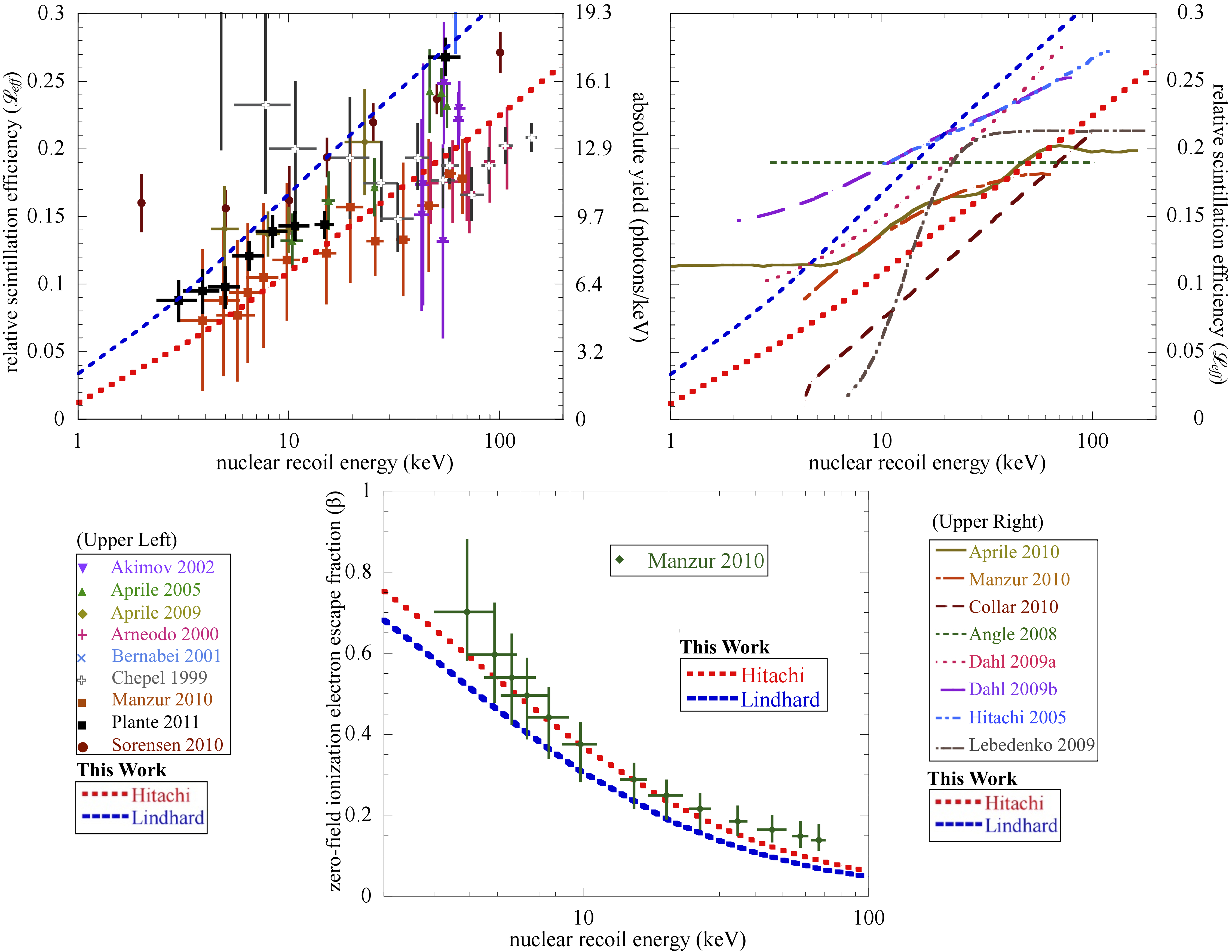}
\caption[]{(top row) Our predictions for NR compared with all past data and predictions. For clarity, we have divided the data sets into  discrete (left) and continuous (right), as presented by the authors. Data are gathered from existing compilations as well as from the primary sources~\cite{Manzur2010,Dahl2009,Collar2010,Plante2011,Sorensen2010,Hitachi2005,Sorensen2009}. Errors bands for data in the right plot have been omitted for further clarity. The axis labeled relative scintillation efficiency ($\mathcal L_{eff}$) refers to a  traditional normalization to $^{57}$Co, a mixture of 122 keV (dominant) and 136 keV gammas. Note that our curves are not fits but predictions. We plot the same predictions at top left and top right. Surprisingly, given the expected inapplicability of the Lindhard theory to liquid xenon, unmodified it appears to agree the best with the latest measurements.  It is in good agreement with Plante\etal\cite{Plante2011} and mostly resides within 1$\upsigma$ of Manzur\etal\cite{Manzur2010}, which seems better described with the Hitachi correction.  (Note: we do not include works that are primarily commentary on older data~\cite{Sorensen2010,Savage2010}, but do include other theoretical and phenomenological predictions, extrapolations, and calculations, like that of Collar~\cite{Collar2010}.)

(bottom) We plot the escape electron fraction for further verification of our model. Since the escape electrons cannot be collected at zero field, the points listed as Manzur\etal are not a direct measurement: they are an extrapolation based on a model Manzur\etal use to provide a good fit to their scintillation yield data, one which included the basic assumption of light-charge anti-correlation~\cite{Manzur2010}.}
\label{NuclearRecoil}
\end{centering}
\end{figure}

We use the same value of $\xi/N_{i}$ ($= 0.19$) as for ER to evaluate the model against empirical data. A choice remains in the application of the Lindhard factor with respect to proposed  corrections. We elect to investigate two scenarios in order to treat this intricate subject with due diligence: pure Lindhard, and the Hitachi correction to the Lindhard factor \cite{Sorensen2011}. Lindhard theory is more appropriate for solid crystal scintillators, and for xenon nuclei Hitachi has performed a first-principles recalculation~\cite{Sorensen2011}. We thus closely follow the methodology of Sorensen and Dahl, but we differ by starting with an electron-recoil-motivated Thomas-Imel parameter at zero electric field instead of one found from fits to NR data~\cite{Sorensen2011}. Therefore, we offer a novel approach.

Our two predictions in Figure~\ref{NuclearRecoil} are consistent with recent measurements by Plante\etal\cite{Plante2011} and Manzur\etal\cite{Manzur2010}, but in direct conflict with the older claims of Sorensen\etal\cite{Sorensen2010} and Aprile\etal\cite{Aprile2010}. Because NR yields change very weakly with field~\cite{Aprile2006, Aprile2005}, we did not initially investigate field dependence. Dahl's study suggests an explanation for this weak dependence: $\alpha \sim$ 1 for NR at non-zero field and perhaps is field-dependent, not a fixed 0.06 as in ER~\cite{Dahl2009}. (Energy dependence at low energy is also possible~\cite{Sorensen2011}). If a significant fraction of interactions is due to excitations, then recombination probability is less important. For zero field we saw no need to change the $\alpha$ or $\xi/N_{i}$ used successfully earlier. To reproduce field dependence, which is of greater importance now as S2 is used to lower the threshold of xenon detectors~\cite{Sorensen2010b}, we suggest employing a $\xi/N_{i}$ that changes differently with electric field than it does for ER, rather than attempting a change in $\alpha$. Though this is in apparent contradiction with Dahl's work, we offer it as another means to the same end. The ability to distinguish between microphysics interpretations (changing $\alpha$ or changing $\xi/N_{i}$) is likely beyond the reach of macroscopic empirical measurements that observe yields.

\section{Conclusions} 
\label{sec:conclusion} 

We have presented a coherent and comprehensive approach towards understanding the scintillation and ionization processes in liquid xenon.  Starting with an ansatz-driven approach, aided by physically motivated models, we have been able to not only explain a large majority of the world data, but make predictions in regions where measurements are sparse. Our model is especially applicable to the low recoil energy regimes of interest for direct dark matter searches, as demonstrated by our prediction of $\mathcal L_{eff}$. NEST will be made available to the world community to be used as an add-on to Geant4. Changing a few parameters in NEST, not all free ($W$, $\alpha$, $A$, etc.), will make it possible to simulate noble elements other than xenon. We have also suggested a universal nomenclature ($W, W_{ex}, W_{i}, W_{ph}, W_{i}^{eff}$) in order to unify different definitions in the literature, which would enable the community to compare and contrast results which are often quoted in varying and confusing fashions. Following in the footsteps of other recent works striving for a unified view \cite{Dahl2009, Sorensen2011, Shutt2007}, we suggest the use of a mean work function (simply labeled as  $W$) as the standard for reporting results and as the benchmark for quoting relative yields. We have provided a framework for simulating electron and nuclear recoils at different energies and applied electric fields, and have confirmed the results against a large sample of data from literature.

\acknowledgments

This work was supported by U.S. Department of Energy grant DE-FG02-91ER40674 at the University of California, Davis as well as performed in part under the auspices of the Department of Energy by Lawrence Livermore National Laboratory under Contract DE-AC52-07NA27344. We would like to thank C. E. Dahl for numerous in-depth discussions and paper draft reviews, but especially for laying the foundation for our paper through his seminal work with T. Shutt and J. Kwong at Case Western Reserve University. We also thank L. Baudis, A. Bolozdynya, L. Kastens, K. Ni, K. Masuda, D.N.~McKinsey, and M. Yamashita for answering questions about their past work and for forwarding work with which we were not familiar or could not easily acquire, and R. Svoboda for consulting on drafts and reviewing them.

\bibliographystyle{JHEP}
\bibliography{XenonPaper}
\end{document}